\documentclass[useAMS,usenatbib]{mn2e}
\usepackage{graphicx}
\usepackage{multirow}
\usepackage[centertags]{amsmath}
\usepackage{color}
\usepackage{amsfonts}
\usepackage{amssymb}

\title[Oscillation modes in the SPB star $\mu$ Eridani]{Oscillation modes
 in the rapidly rotating Slowly Pulsating B-type star $\bmu$ Eridani}
\author[Daszy\'nska-Daszkiewicz et al.]{J. Daszy\'nska-Daszkiewicz$^{1}$\thanks{E-mail:
daszynska@astro.uni.wroc.pl}, W. A. Dziembowski$^{2,3}$, M. Jerzykiewicz$^1$ and G. Handler$^3$\\
$^{1}$Instytut Astronomiczny, Uniwersytet Wroc{\l}awski, Wroc{\l}aw, Poland \\
$^{2}$Warsaw University Observatory, Al. Ujazdowskie 4, 00-478 Warsaw, Poland\\
$^{3}$Copernicus Astronomical Center, ul. Bartycka 18, 00-716 Warsaw, Poland
}

\begin{document}

\date{}

\pagerange{\pageref{firstpage}--\pageref{lastpage}} \pubyear{}

\maketitle

\label{firstpage}

\begin{abstract}
We present results of a search for identification of modes
responsible for the six most significant frequency peaks detected in the
rapidly rotating SPB star $\mu$ Eridani. All published and some
unpublished photometric data are used in our new analysis.
The mode identification is carried out with the method
developed by Daszy\'nska-Daszkiewicz et al. employing the phases and
amplitudes from multi-band photometric data and relying on the
traditional approximation for the treatment of oscillations in rotating stars.
 Models consistent with the observed mean parameters are
considered. For the five frequency peaks, the candidates for the
identifications are searched amongst unstable modes. In the case of
the third frequency, which is an exact multiple of the orbital
frequency, this condition is relaxed. The systematic search is
continued up to a harmonic degree $\ell =6$.
 Determination of the angular numbers, $(\ell,m)$, is done
simultaneously with the rotation rate, $V_{\rm rot}$, and the
inclination angle, $i$, constrained by the spectroscopic data on the
projected rotational velocity, $V_{\rm rot}\sin i$, which is assumed
constant. All the peaks may be accounted for with g-modes of high
radial orders and the degrees $\ell\le 6$. There are differences in
some identifications between the models. For the two
lowest--amplitude peaks the identifications are not unique.
Nonetheless, the equatorial velocity is constrained to a narrow
range of (135, 140) km/s.
 Our work presents the first application of the photometric method
of mode identification in the framework of the traditional
approximation and we believe that it opens a new promising direction
in studies of SPB stars.
\end{abstract}


\section{Introduction}

Variability with a few periods of the order of one day is frequently observed in
main-sequence B-type stars \citep[e.g.][]{DeC07}. Objects exhibiting only these slow
oscillations are now called {\bf S}lowly {\bf P}ulsating {\bf B}-type stars (SPB) and
are found in the spectral-type range of B3-B8. It is commonly believed that this
variability is due to excitation of high-order g-modes. Such modes are also invoked as
an explanation of low-frequency variations observed in certain $\beta$ Cephei stars.
Linear non-adiabatic calculations predict excitation of high-order g-modes in a wide
range of main sequence B-type stars. However, difficulties in explaining in this way
long-period variability in certain objects have been noted.

In several stars, observations from space by the CoRoT and {\it Kepler} missions led to
the detection of a large number of frequency peaks which may be associated with g-modes.
However, mode identification for individual peaks seems a formidable task. A method
based on a search for equidistant spacing in period, that has been successfully used in
the application to white dwarfs, was tried for the CoRoT target HD50230 by
\citet{DAB+10}. The authors found a sequence of eight modes with a remarkably constant
separation, much more so than in relevant models \citep*{DMP93, AD12}. In the same
frequency range there were over 100 frequencies (some of much higher S/N than most of
those forming the sequence) for which no plausible interpretation was offered so far
\citep{SDD14}. In an attempt to explain the oscillation spectrum of another CoRoT
target, HD 43317, \citet{S13} also could not account for the equidistant spacing reported
by \citet{PBB+12}. Patterns in oscillation spectra apparently do not yield a key toward
deciphering information on excited modes in SPB stars. We need to know more about mode
selection in these stars. Perhaps we may learn something new about it from ground-based
observations, which yield much fewer frequency peaks than the ones from space, but much more
information about each of them that would enable mode identification. In this respect,
the SPB star $\mu$ Eri appears to be the best choice.

A large number of three-passband photometric data for this star have been collected
during two multisite campaigns (MSC) aimed at the $\beta$ Cep variable $\nu$ Eri,
2002-2003 MSC \citep{HSJ+04} and 2003-2004 MSC \citep[][hereafter J2005]{jhs}. The star
has been used as one of two comparison stars in both campaigns. The analysis of the
2002-2003 MSC data by \citet{HSJ+04} revealed a variability of $\mu$ Eri with a
frequency of 0.616 c/d. However, the authors were not able to decide what is the origin
of this variation: pulsation or rotational modulation. The analysis of the extensive
data set from both campaigns presented in J2005 led to the detection of five additional
significant peaks in the g-mode frequency range typical for SPB stars and an unambiguous
classification of $\mu$ Eri.

Additional three-colour data were recently obtained by one of us (GH).
These and the MSC data were employed in the mode identification presented in our work.
Photometry with the MOST micro-satellite \citep[][hereafter J2013]{J+13}
was also used in frequency determination in Section\,2.3. Unfortunately, spectroscopic
observations reported in the same paper yielded only the upper bound on radial velocity
variations. However, these observations were used to constrain parameters of the star;
this, as we will see later in the paper, is essential for mode identification.

All data used in our work are presented in the next section. We also provide information
on the mean physical parameters of the star and describe its models which will be used
in the subsequent sections.

In Section\,3, we confront frequencies of the peaks with instability ranges determined
in one of the selected models. We use the condition of mode instability as a requirement
for its possible association with an observed frequency. Mode visibility is another
factor limiting the set of modes for possible identification. This matter is discussed
in Section\,4.

Section\,5 is devoted to an analysis of the data on the amplitudes and phases of the six
peaks determined in three photometric passbands. The aim is to determine the harmonic
degree, $\ell$, and the azimuthal order, $m$, of the observed modes simultaneously with
constraints on the stellar rotation rate, assuming that it is uniform. The method, which
employs the traditional approximation, has been described in detail by
\citet*[][hereafter D07a]{DDP07a} and here we give only its brief outline.

We have not attempted to constrain a stellar model by fitting calculated frequencies to
observations because it would not make sense in view of the approximation adopted in
treating the effects of rotation and high density of the calculated oscillation spectra
related to high radial orders of the excited modes. Our primary goal is to learn
something about mode selection in SPB stars. We believe that this is essential for
understanding complicated oscillation spectra from the space missions. The discussion in
the final section is focused on this issue.

\section{$\bmu$ Eridani}

$\mu$ Eridani (HD 30211) is a bright ($V ={}$4 mag) B5 IV star. It is certainly not a
slow rotator. J2013 provide two independent estimates of the projected equatorial
velocity of rotation, $V_{\rm rot}\sin i ={}$130$\,\pm\,$3 and   136$\,\pm\,$6 km/s.  It
has been known since a long time \citep{FL10} that the star is the primary in a
single-lined spectroscopic binary system but only J2005 detected shallow eclipses. This
is a well detached system with the orbital period $P_{\rm orb} ={}$7.38090 d and
eccentricity $e ={}$0.344 (J2013). Hence, we can neither assume synchronized rotation
nor parallel rotation and orbital axes. Thus, mode identification must be done
simultaneously with the determination of the inclination angle.

\subsection{Mean parameters and evolutionary models}

We adopted after J2013 $\log T_{\rm eff} ={}$4.195$\,\pm\,$0.013 and $\log L/{\rm
L}_{\sun} ={}$3.280$\,\pm\,$0.040 for the effective temperature and luminosity of $\mu$
Eri. These values were found from the mean Str\"omgren photometric indexes  and the
revised value of the {\it Hipparcos} parallax \citep{vL}. From their own spectroscopic
data, J2013 found a metal abundance close to solar. They also determined the mean
density of the star, $\bar\rho ={}$0.0348$\,\pm\,$0.0032 g/cm$^3$ from the orbital
parameters.

The above-mentioned values of $\log T_{\rm eff}$, $\log L/{\rm
L}_{\sun}$ and $\bar\rho$ imply a mass $M \approx{}$6.0 M$_{\sun}$
and a radius $R \approx{}$6.1 R$_{\sun}$, hence the break-up
velocity amounts to $V_{\rm rot}^{\rm crit}\approx
\sqrt{\frac{GM}{R}} \approx{}$ 430 km/s and the corresponding value
of the critical rotational frequency to $\nu_{\rm rot}^{\rm crit}
\approx{}$1.3 c/d. Thus, $\mu$ Eri rotates at a speed of at least
30\% of $V_{\rm rot}^{\rm crit}$. Our models are spherically
symmetric and include only the spherically symmetric effect of
centrifugal force. This is a crude but an acceptable approximation
if $(\Omega/\Omega_{\rm crit})^2 \lesssim 0.5$, where $\Omega$ and
$\Omega_{\rm crit}$ are the angular frequency of rotation and its
critical value, respectively. Assuming for $\mu$ Eri the radius of
$R\approx 6.2R_{\sun}$, we obtain $(\Omega/\Omega_{\rm
crit})^2\approx 0.5$ for $V_{\rm rot}=280$ km/s and this number is
the upper limit in our search for the mode identification. The
adopted lower limit is 130 km/s and it results from the value of
$V_{\rm rot}\sin i=130$ km/s. In all our models, rotation is assumed
uniform.

All the stellar models in this work were calculated with the Warsaw evolution code
adopting OPAL opacities \citep{IR96} for  $Y$ =0.28, $Z$=0.015 and the AGSS09 solar
mixture of heavy elements \citep{asp09}. Parameters of the models adopted for mode
identification in $\mu$ Eri are listed in Table\,1. As may be seen in Fig.\,1, where
positions of the models on the evolutionary tracks in the HR diagram are shown, these
models satisfy the observational constraints discussed above. At a finite rotation rate,
positions in the HR diagram depend on $i$. What is shown are aspect-averaged positions
but the spread is small in comparison with the mean shift which we may asses by
comparing tracks with the initial equatorial velocity of 200 and 0 km/s.

The three selected models were calculated adopting $V_{\rm rot} ={}$200 km/s at the ZAMS
and the overshooting parameter, $\alpha_{\rm ov} ={}$0.2. Note in Fig.\,1 that the track
with $\alpha_{\rm ov} ={}$0 enters the error box but only at the edge of the strip allowed
by the mean density. Model 2 is on the same track as Model\,1 but it is more evolved.
Model 3 has a similar temperature to Model 2 but has seven percent higher surface
gravity and is less evolved. All these models lie within the main sequence, close to its
red boundary (TAMS).
\begin{table}
\caption{Models considered for $\mu$ Eridani.}
\begin{tabular}{ccccccc}
\hline
 Model & $M$        & $V_{\rm rot}$ & $\alpha_{\rm ov}$& $\log T_{\rm eff}$ &  $\log L$   & $\nu_{\rm rot}$  \\
       & [M$_{\sun}]$&  [km/s]       &     [$H_p$]      &         [K]        & [L$_{\sun}$] &     [c/d] \\
\hline
1 & 6.00 & 170 & 0.2 & 4.1939 & 3.300 & 0.55\\
2 & 6.00 & 167 & 0.2 & 4.1861 & 3.311 & 0.51 \\
3 & 5.85 & 166 & 0.2 & 4.1848 & 3.264 & 0.53 \\

 \hline
\end{tabular}
\end{table}
\begin{figure}
\includegraphics[width=\columnwidth,clip]{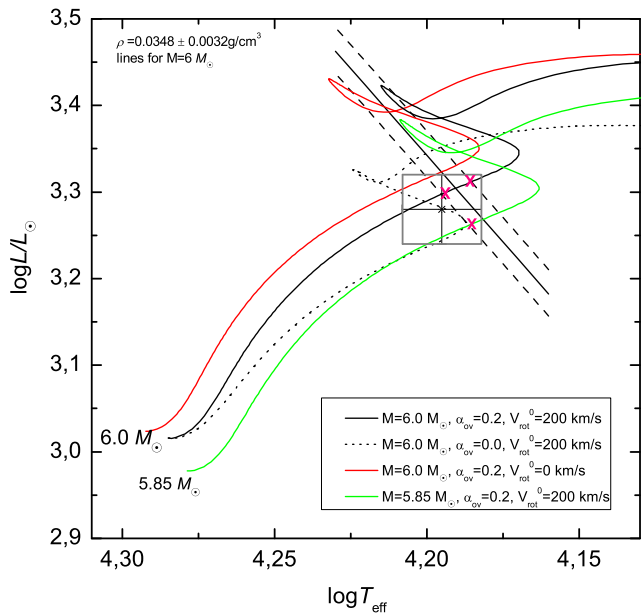}
\caption{ The error box from J2013 shows the position of $\mu$ Eri in the HR diagrams.
The oblique straight lines correspond to the range of the mean density found in J2013 at
$M ={}$6 M$_{\sun}$. The evolutionary tracks shown in this figure were calculated at
metallicity $Z ={}$0.015. Values of mass, overshooting parameter, $\alpha_{\rm ov}$, and
the equatorial velocity, $V_{\rm rot}$, at the ZAMS are given in the legend. Selected
models are shown with the crosses.}
\end{figure}

\subsection{New three-band photometric data}

New photometric measurements of $\mu$ Eri, consisting of differential time-series
photoelectric data in the Str\"omgren $uvy$ filters, were acquired in the season
2012-2013. The star was measured alternatingly with the $\beta$ Cephei star $\nu$ Eri
and $\xi$ Eri, using the 0.75-m Automatic Photoelectric Telescope (APT) T6 at Fairborn
Observatory in Arizona.

The data were reduced in a standard way for photoelectric time-series photometry. First,
the measurements were corrected for coincidence losses. Then, sky background was
subtracted and nightly extinction coefficients, determined from the measurements of
$\xi$ Eri via the Bouguer method (fitting a straight line to a magnitude {\it vs}\
air-mass plot). At a given air mass, the same extinction correction was applied to each
star. Finally, differential magnitudes were computed by interpolation, and the timings
were converted to Heliocentric Julian Date. We ended up with some 1120 data points for
$\mu$ Eri obtained in 91 nights, for a total of 365 hr of measurement with a time span
of 187.8 d.

We note that \citet{HSJ+04} and J2005 reported low-amplitude variability of $\xi$ Eri,
probably due to a $\delta$ Scuti-type pulsation. As such variations could not be
detected in the present data set, we treated $\xi$ Eri as a constant comparison star.
The time series obtained for $\nu$ Eri will be discussed elsewhere; we concentrate on
$\mu$ Eri in what follows.

\subsection{Significant frequency peaks}

For the frequency analysis, we used the 2002-2003 and 2003-2004 MSC
$v$-filter data, the 2009 MOST data, and the 2012-2013 APT
$v$-filter data. In all cases, observations between the first and
the last contact of the eclipses, i.e. those falling within
$\pm\,$0.03 orbital phase away from the mid-eclipse epochs computed
from the ephemeris provided by J2013, were omitted. In order to
eliminate the short-period variations of unknown origin in the MOST
data, the data were averaged in adjacent segments of about 0.044 d
duration (see the Appendix in J2013). For consistency's sake, the
ground-based data were treated likewise. Mean light-levels were
subtracted from the ground-based data; the MOST data were de-trended
(J2013 Section\,3.1). After these modifications, the data included
1354 data-points (845 2002-2003 and 2003-2004 MSC, 204 MOST, and 305
2012 APT), spanning 3850 d; the frequency resolution was therefore
equal to 0.00026 c/d. The analysis consisted of computing  an
amplitude spectrum, identifying the highest peak, pre-whitening etc.
The results were the following (in the parentheses, the
signal-to-noise ratio is given): 0.6158 c/d (14.4), 0.7009 c/d
(8.9), 1.2056 c/d (6.6), 0.6589 c/d (5.4), 0.5681 c/d (5.1), 0.8129
c/d (4.4), 0.4046 c/d  (4.1), and 1.1810 c/d (3.8). The noise was
computed as a straight mean of the amplitudes in the frequency range
0.3 to 1.3 c/d, obtained from the amplitude spectrum of the data
pre-whitened with the above eight frequencies.

The first six frequencies are very nearly equal to the frequencies
$f'_1$, $f'_2$, $f'_4$, $f'_5$, $f'_6$, and $f'_3$ of table 7 of
J2005. In all cases the signal to noise is greater than 4, so that
they all fulfill the popular criterion of significance of
\citet{B+93}. Additional arguments in favour of the significance of
$f'_5$, $f'_6$, and $f'_3$ are the following (1) the frequency
quadruplet $f'_6$, $f'_1$, $f'_5$, $f'_2$, although not resolved by
the MOST data, is necessary to account for the MOST doublet $f_1$,
$f_2$ (see J2013, Section 9.1) and (2) $f'_3$ is close to the highly
significant MOST frequency $f_7$ (the difference $f_7 - f'_3$ is
equal to one fourth of the frequency resolution of the MOST data).
Thus, we retained the six frequencies, leaving the remaining ones
for future verification by means of satellite data with sufficient
frequency resolution. Using the frequencies $f'_i$ ($i ={}$1,..,6)
as starting values in a nonlinear least-squares fit to the
ground-based not-averaged $v$ data we obtained: $\nu_1
={}$0.6157633$\,\pm\,$0.0000021, $\nu_2
={}$0.7008807$\,\pm\,$0.0000044, $\nu_4
={}$1.2055398$\,\pm\,$0.0000046, $\nu_5 ={}$0.6588788$\,\pm\,$
0.0000058, $\nu_3 ={}$0.812899$\,\pm\,$0.000008 and $\nu_6
={}$0.568107$\,\pm\,$0.000007 c/d. Finally, for $\nu_i$ ($i
={}$1,..,6), the $uvy$ amplitudes and phases were computed by means
of linear least-squares using the ground-based data. The results are
listed in Table 2. The standard deviations of the amplitudes and
phases (given in parentheses without the leading zeroes) are the
formal standard deviations of the least-squares solutions multiplied
by two. In this, we follow \citet{H+00} and J2005 who---while
dealing with time-series observations similar to the present
ones---showed that the formal standard deviations were
underestimated by a factor of about two.
\begin{table}
\caption{Amplitudes and phases in the $uvy$ Str\"omgren passbands
for six frequency peaks found in the light curve of $\mu$ Eri.
The function $\sum_{i=1}^6 A_i\cos(2\pi\nu_i (t-t_0)-\varphi_i)$ was
fitted with the zero-epoch, $t_0$, equal to HJD 245 4450.}
\begin{tabular}{ccccccc}
\hline
        & $A$ [mag] & $\varphi$ [rad]  \\
\hline
\multicolumn{3}{c}{$\nu_1 ={}$0.6157633(21) c/d, $S/N  ={}$14.4} \\
\hline
$u$ &   0.00974(15) &  4.2832(148)\\
$v$ &   0.00626(11) &  4.3011(169)\\
$y$ &   0.00536(10) &  4.3342(177)\\
\hline
\multicolumn{3}{c}{$\nu_2 ={}$0.7008807(44) c/d, $S/N  ={}$8.9} \\
\hline
$u$ &   0.00467(14) &  3.9164(308)\\
$v$ &   0.00292(11) &  4.0975(361)\\
$y$ &   0.00238(10) &  4.0058(s397)\\
\hline
\multicolumn{3}{c}{$\nu_3 ={}$0.812899(8) c/d, $S/N  ={}$4.4} \\
\hline
$u$ &   0.00224(14) &  1.6037(647)\\
$v$ &   0.00168(11) &  1.7779(634)\\
$y$ &   0.00115(10) &  2.0308(820)\\
\hline
\multicolumn{3}{c}{$\nu_4 ={}$1.2055398(46) c/d, $S/N  ={}$6.6} \\
\hline
$u$ &   0.00327(15) &  1.7280(436)\\
$v$ &   0.00275(11) &  1.9930(384)\\
$y$ &   0.00251(10) &  2.0943(378)\\
\hline
\multicolumn{3}{c}{$\nu_5 ={}$0.6588788(58) c/d, $S/N  ={}$5.4} \\
\hline
$u$ &   0.00375(14) &  1.4119(389)\\
$v$ &   0.00235(11) &  1.4735(460)\\
$y$ &   0.00234(9)  &  1.4820(416)\\
\hline
\multicolumn{3}{c}{$\nu_6 ={}$0.568107(7) c/d, $S/N  ={}$5.1} \\
\hline
$u$ &   0.00264(14) &  1.8346(562)\\
$v$ &   0.00183(11) &  1.9181(599)\\
$y$ &   0.00176(9)  &  1.8882(549)\\
\hline
\end{tabular}
\end{table}

 The frequency $\nu_3$ is an exact multiple of the orbital
frequency, i.e., $\nu_3=6/P_{\rm orb}$. Given the significant
eccentricity of the binary system, this frequency may arise as a
tidal effect. As we shall see in Section\,5, assuming different
origins of this frequency requires only a small modification of our
search for mode identification and it is inconsequential for the
results.

In Fig.\,2, we show the amplitude ratios in two colours plotted against
the corresponding phase differences. Such diagrams are often used for
revealing the mode degree, $\ell$. If the effects of rotation are
large, a direct inference is not possible. Nonetheless, the diagrams are still
useful for presenting data and their errors. Two pairs of passbands
were considered: $(u,v)$ and $(u,y)$. Let us note the large phase
differences for peaks $\nu_3$ and $\nu_4$.

Typically for SPB stars the phase difference in any two passbands is
close to zero, which is expected if the temperature variations
dominate. Ignoring the effects of rotation, \citet{T02} and
\citet{DeC+04} found the $\ell ={}$1 identification for most modes
in SPB stars known at that time.  However, most of these SPB stars
are slower rotators than $\mu$ Eri as shown by \citet{DeC+05}. Let
us remind the reader that a dipole mode does not cause variations of
the star's shape.

\begin{figure*}
\centering
\includegraphics[width=\textwidth, clip]{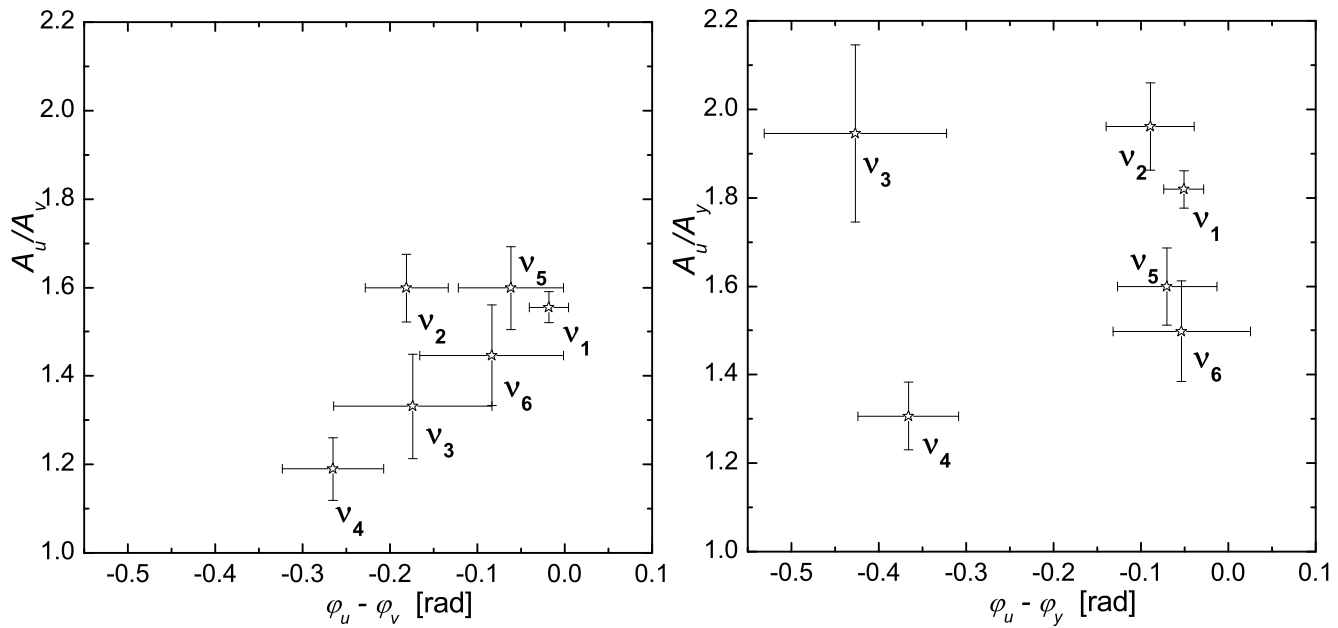}
\caption{Amplitude ratios and phase differences for significant frequency peaks of $\mu$ Eri
in the photometric diagnostic diagrams involving two pairs of the Str\"omgren passbands:
$(u,v)$ and $(u,y)$, showed in the left and right panel, respectively.}
\end{figure*}

\section{Unstable modes}

Looking for possible identifications of the frequency peaks, we
focus on unstable modes.  In Section\,5, we will abandon this
requirement for the frequency $\nu_3$ which may be resonantly
excited by the companion to the pulsator in the $\mu$ Eri binary
system.

At these low frequencies, we must consider not only gravity modes
but also mixed-Rossby modes, which appear at high
rotation rate. They are retrograde with $m=-\ell$. In the present
paper we will refer to them as the r-mode and identify them through the
$m$ value alone. \citet{S05} and \citet{T05} showed that they may
become unstable. In our stability analysis we use a nonadiabatic
generalisation of the traditional approximation (\citep*{DzDP07}).
The only modification of the equations for nonadiabatic oscillations
in nonrotating  stars (in the Cowling approximation) is the
replacement of $\ell(\ell+1)$ with $\lambda$, which is the
eigenvalue of the Laplace tidal equation \citep[see e.g.][]{BUC96,
LS97, T03a}. At finite rotation rate, we have
$\lambda=\lambda(\ell,m,s)$, where $s=2\nu_{\rm rot}/\nu_*$ (the
spin parameter) and $\nu_*$ is the eigenmode frequency in the
corotating system.  For all g-modes
$\lambda(\ell,m,0)=\ell(\ell+1)$. For the prograde sectorial modes
($m=\ell$), $\lambda$  slowly decreases with $s$,
asymptotically tending to $m^2$, whereas for other modes it sharply
increases \citep*[e.g.][]{T03a}, which is also true for the r-modes
once they appear at $s=|m|+1$. We put the azimuthal and temporal
dependence in the form ${\rm exp}[{\rm i}(m\varphi-2\pi\nu_* t)]$,
which implies $m>0$ for prograde modes and $m<0$ for retrograde
modes. Thus, the frequency in the inertial system is given by
$\nu=|\nu_*+m\nu_{\rm rot}|$.

\begin{figure}
\includegraphics[width=\columnwidth,clip]{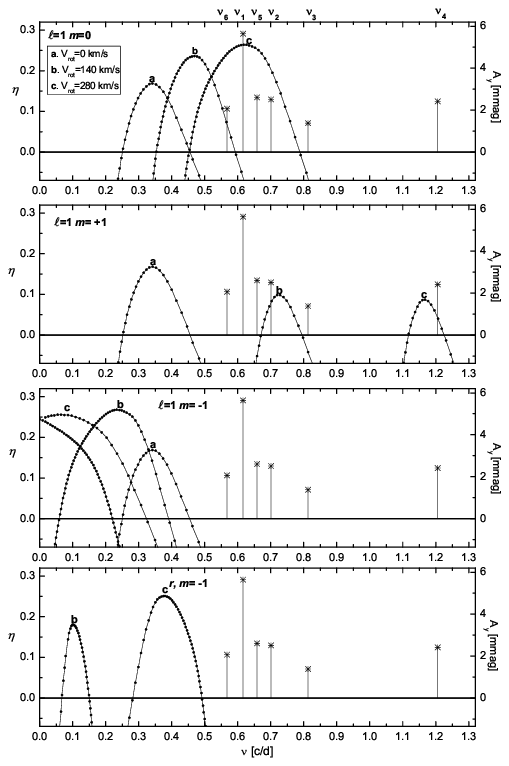}
\caption{ The normalized instability parameter $\eta$ as a function of the frequency in
the inertial system for g-modes at $\ell ={}$1 and r-modes at $m ={}-$1 in Model 1 (see Table\,1).
Three values of a rotational velocity were considered: 0, 140 and 280 km/s, marked
as {\bf a, b, c}, respectively. Amplitudes of the observed peaks in the $y$ Str\"omgren
passband are given on the right axis. }
\end{figure}
\begin{figure}
\centering
\includegraphics[width=\columnwidth,clip]{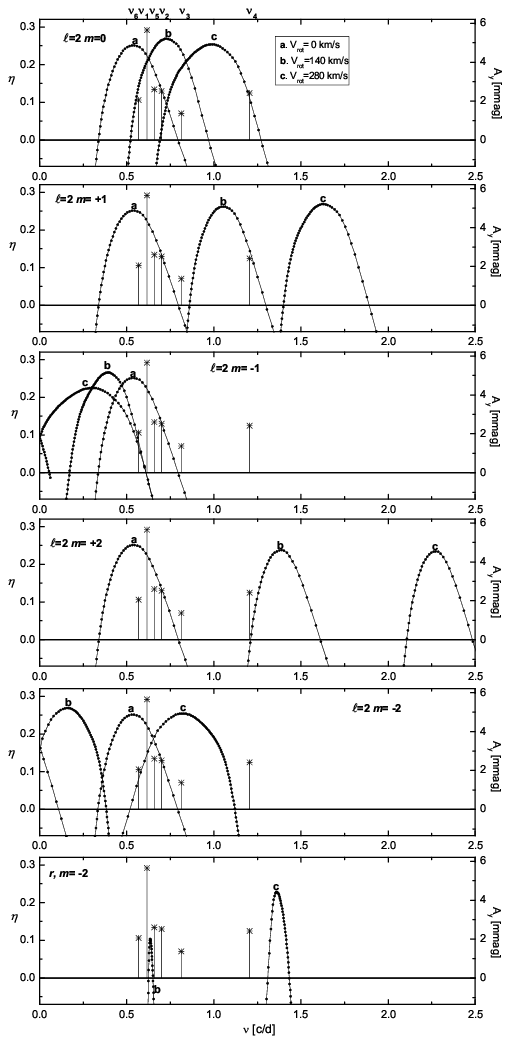}
\caption{ The same as Fig.\,3 but for g-modes at $\ell ={}$2 and
r-modes at $m ={}-$2.}
\end{figure}

According to \citet{L10}, the number of unstable retrograde g-modes is much smaller if
the truncated Legendre expansion is used instead of the traditional approximation. Thus,
perhaps  some of the modes that we allow should be excluded. On the other hand,
\citet{S07} with his 2D models largely confirmed results of \citet{DzDP07} regarding
instability of the retrograde modes obtained with the approximation adopted in the
present work.

Following \citet{S78}, we measure mode instability with the normalized growth rate
$$\eta=\frac{W}{\int_0^R \left|\frac{dW}{dr}\right| dr}$$
where $W$ is the usual work integral over the pulsational cycle. From this definition, it
follows that $\eta$ varies in the range $[-1,+1]$ and it is positive ($\eta>0$) for
unstable modes, that we regard as possible candidates for interpretation of the
oscillation spectrum of $\mu$ Eri.

In Fig.\,3, we show the values of $\eta$ plotted against frequency
in the observer's system for gravity modes with $\ell ={}$1 and
r-modes with $m ={}-$1 in Model\,1. In each panel we show also
frequencies and the $y$ Str\"omgren amplitudes of the six
significant peaks detected in $\mu$ Eri. The mode frequencies were
calculated for rotation rates 0, 140 and 280 km/s. Let us remind
that in this model the mean effect of centrifugal force is evaluated
at $V_{\rm rot}\approx{}$170 km/s. However, our intention here was
to show only the effects of rotation induced by the Coriolis force
and the Doppler effect. Consequences of changes in the model caused
by the changes of centrifugal force at the same $T_{\rm eff}$ are
much smaller. Fig.\,4 shows the same but for the $\ell ={}$2 g-modes
and the $m ={}-$2 r-mode.

The top panels of Figs.\,3 and 4 show a pure effect of the Coriolis force ($m ={}$0).
The frequencies of the unstable modes increase with rotation rate. In the adopted
approximation, the effect is equivalent to a continuous increase of $\ell$ in
nonrotating stars. Notice that the sequence ,,b'' in Fig.\,4 is quite similar to the
sequence ,,c'' in Fig.\,3. Both cover the range of the four lowest frequency peaks.
However, these peaks cannot be associated with consecutive radial orders, as thought by
J2005, because the separations between them are too large.  If rotation is fast enough,
the $\ell ={}$1, $m ={}$0 and $\ell ={}$2, $m ={}$0 identifications are also possible
for $\nu_3$. The latter identification is also acceptable for $\nu_4$.

The plots for the $\ell=m$ modes reveal mainly the effect of the
Doppler frequency shift because changes of $\lambda$ for sectoral
prograde modes are small. Since 130 km/s is adopted as the minimum
equatorial velocity, the $m=\ell ={}$1 identification could be the only one
for the four highest frequency peaks and the $m=\ell ={}$2 identification,
the only one for the $\nu_4$ peak.

None of the peaks may be associated with the $\ell ={}$1, $m ={}-$1
mode nor the $m ={}-$1, r-mode. The $\ell ={}$2, $m ={}$1
identification is possible for $\nu_4$ but at a lower rotation rate
than the $\ell ={}$2, $m ={}$0 identifications. The $\ell ={}$2, $m=
-{}$1 mode may be responsible for one or two peaks of lowest
frequency. The $\ell ={}$2, $m ={}-$2 identification is possible for
all the peaks except for $\nu_4$ if rotation is fast enough. The
instability ranges for the r-modes are narrower than for g-modes.
Yet, at $m ={}-$2 the r-modes may account for all observed peaks
except $\nu_6$.

Instability of g-modes continues well beyond $\ell ={}$2. In Model\,1
it goes up to $\ell ={}$17 and extends for many radial orders.
At the zero rotation rate there are altogether about $1.5\times10^4$
unstable modes. At $V_{\rm rot} ={}$280 km/s, the number of unstable
modes is about 30 percent lower. The effect of increasing rotation rate
on the r-mode instability is opposite but the number of unstable r-modes
is always much smaller than that of g-modes.

The large number of unstable modes follows, in part, from their high
radial orders, $n$, which range from 25 to 101 and imply small
frequency separations between consecutive modes, and, in part, from
the favorable shape of the eigenfunctions (relatively large amplitudes
in the driving zone) in wide frequency ranges.

\begin{figure}
\includegraphics[width=\columnwidth, clip]{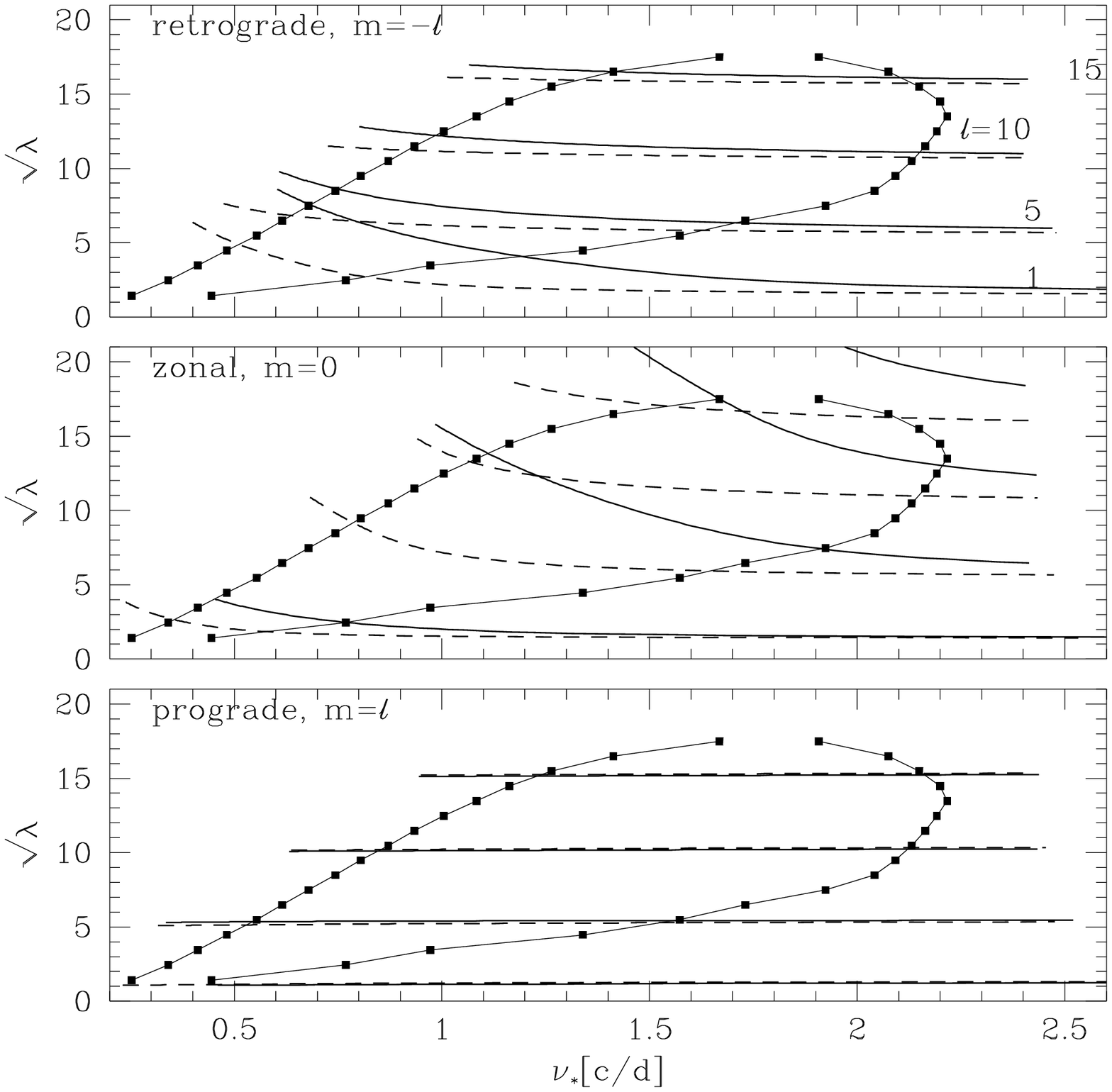}\caption
{Instability domain in the $(\lambda^{1/2}, \nu_*)$ plane for
Model\,1. Pairs of symbols at the same values of
$\lambda=\ell(\ell+1)$ determine the frequency range of unstable
modes at $\ell ={}$1 to 17. These ranges were calculated with the
standard nonadiabatic pulsation code ignoring effects of the
Coriolis force. The instability domain is the same  with the
traditional approximation but then $\lambda$ depends on $\nu_*$. In
the three panels, examples of the $\lambda^{1/2}(\nu_*)$ dependence
for retrograde, zonal, and prograde modes are shown. The solid and dashed
lines correspond to $V_{\rm rot} ={}$280 and 140 km/s, respectively.}
\end{figure}

The mode order and the amplitude behavior in the stellar interior is
determined by the value of $\lambda^{1/2}/\nu_*$. The extent of the
instability range in $\nu_*$ follows from a requirement of a match
between the pulsation period and the thermal time scale in the
driving zone. This sets the upper limit on $\lambda$, which at the
specified $\ell$ and $m$ depends on $V_{\rm rot}$. However, as one
may see in Fig.\,5,  the dependence is weak, except for the zonal
modes. In the two upper panels of Fig.\,5, we may see that at a
specified frequency, $\lambda$ increases with the rotation rate,
causing some modes to move out of the instability domain. Only at
$\ell=m$ the value of $\lambda$ slightly decreases with $V_{\rm
rot}$. For other prograde modes, like for the modes shown in the two
upper panels, it does increase.

The instability ranges are sensitive to the effective temperature.
With the lower $T_{\rm eff}$ and lower $L/M$ ratio, the ranges move to
higher frequencies. Thus, the unstable $\nu_*$- ranges in Model\,2 are shifted
towards the higher frequencies relative to those in Model\,1 and to the lower ones
relative to those in Model\,3.

\section{Visibility of the pulsational modes}

Looking for possible mode identifications, we have to take into account mode visibility.
Large effects of rotation on the slow modes visibility arising at moderate rate were
first studied by \citet{T03b} with the traditional approximation. This approximation was
also adopted in D07a and is used in this paper. Recently, a treatment of the visibility
based on 2D stellar models was developed by \citet{RPB+12} and \citet{S13} with all
dynamical effects of fast rotation taken into account. However, utility of the treatment
presented in the first paper is limited due to the adopted adiabatic relation between
pressure and temperature changes.

In the traditional approximation, the latitudinal dependence of the the radial
displacement is described by the Hough functions, $\Theta_\ell^m(\theta)$, which become
the associated Legendre functions, $P_\ell^m(\theta)$, in the limit of zero-rotation
rate. At finite rotation, mode visibility depends not only on $\ell$, $m$, but also on
the aspect angle, $i$, and the spin parameter, $s$. Our inferred values of $s$ for the
six peaks listed in Table\,2 are between 0.7 and 1.6.

Most of the unstable modes have frequencies much higher than those detected in $\mu$
Eri. Still, if the instability is the only constraint on the mode geometry, the number
of potential identifications to consider is quite large. At $V_{\rm rot} ={}$140 km/s,
there are about 60 possible identifications for each peak with the angular degrees
ranging up to $\ell ={}$17. At $V_{\rm rot} ={}$280 km/s, the number of possible
identifications is reduced to about 20 and the maximum $\ell$ is 14. At high values of
$\ell$, only retrograde modes of moderate $m$ fall into the observed range and, as we
noted in the previous section, they actually may be stable. In any case, the number of
possible identifications remains large if the visibility argument is not taken into
account.

At slow rotation, the decline of the disc averaging factor, $b_\ell$ \citep[see
e.g.][]{DDPG02}, is often used to justify limiting the search for identifications to
modes with $\ell\le2$. Rotation complicates the matter of mode visibility considerably.
The counterpart of this factor within the traditional approximation is given by
$$b_{\ell,m}^x=\int_0^1 h^x(\mu) \mu \Theta_\ell^m (\theta) d\mu, $$
where $\mu=\cos\theta$, $x$ denotes the photometric passband, $h^x(\mu)$ is the
limb-darkening law and $\Theta_\ell^m(\theta)$ are the Hough functions. Now $b_{\ell,m}$
depends additionally  on $m$ and the spin parameter, $s$.

\begin{figure}
\includegraphics[width=\columnwidth, clip]{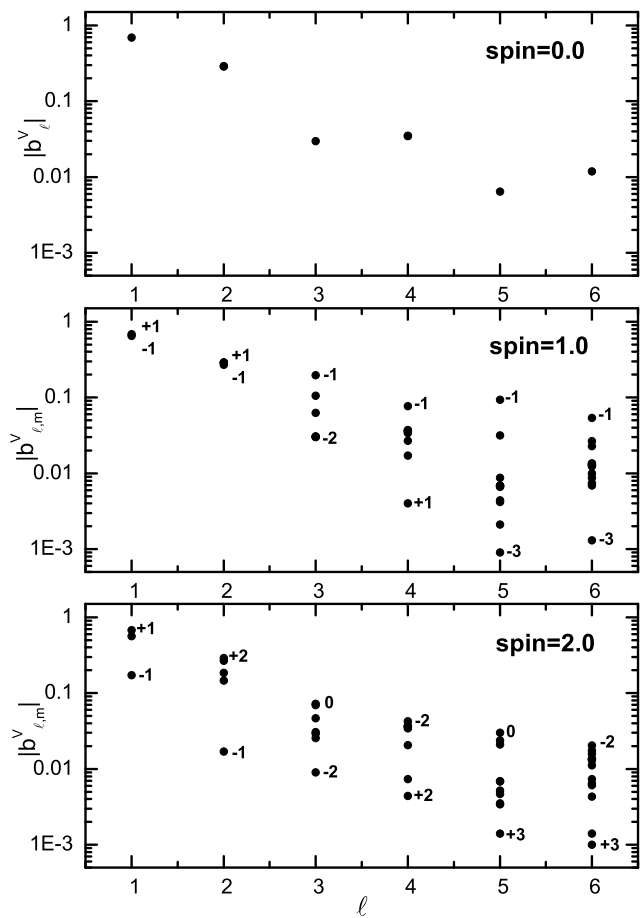}
\caption{The values of the disc averaging factor, $b_{\ell ,m}^V$, in the Geneva $V$
filter as a function of the mode degree, $\ell$, for the three values of the spin
parameter, $s ={}$0.0,1.0,2.0 from top to bottom. The $m$ values at the points
corresponding to extreme values of $b_{\ell ,m}^V$ at each $\ell$ are given. }
\end{figure}

In Fig.\,6, we show the absolute values of $b_{\ell,m}$ in the Geneva $V$ passband as a
function of $\ell$ for three values of $s$. The top panel is for $s ={}$0, i.e., for the
zero-rotation limit. Note the one order of magnitude drop between $\ell ={}$2 and 3. This
drop is significantly reduced to about a factor of 2 at $s ={}$1 and $m ={}-$1. Modes with
$m ={}-$1 at higher values of $\ell$ also do not suffer such reduced visibility. The cause
is a significant contribution from the $P_\ell^{-1}$ component in the expansion of
$\Theta_\ell^m$ in a series of the associated Legendre functions. The situation is quite
different at $s ={}$2. In particular, modes with different azimuthal orders become best
visible. The non-monotonic character of the $b_{\ell,m}(s)$ dependence arises from
changing the contributions from the low- to high-degree $P_\ell^m$ components in
$\Theta_\ell^m$. Thus, the effects of rotation complicate the matter of mode visibility.
In the present work, the systematic search for possible mode identifications for $\mu$
Eri was stopped at $\ell ={}$6.

\section{Modes excited in $\bmu$ Eri and its rotation rate}

The source of observational information on the angular numbers of the excited modes are
the amplitudes and phases in various photometric passbands and in the radial velocity.
Diagrams showing amplitude ratios {\it vs} phase differences, such as those shown in
Fig.\,2, are often used for this goal. If the effects of rotation are negligible, these
parameters are independent of the inclination angle and the azimuthal order, $m$, and,
in favorable conditions, the $\ell$ degree may be inferred from the peak location in the
amplitude ratio {\it vs} phase difference diagnostic diagram.

Mode locations in the diagnostic diagrams at non-zero rotation rate depend on the
inclination angle, azimuthal order and rotational velocity, $V_{\rm rot}$. Even at low
rotation rate, the coupling of modes with the same $m$ and $\ell$ differing by two --
caused by the centrifugal distortion -- may result in large shifts in the diagnostic
diagrams \citep{DDPG02}. This effect is ignored in the traditional approximation adopted
in the present work but it is unlikely to be important for the considered modes.

Here we follow the approach outlined in D07a.

\subsection{The mode discriminant}

The basis of our method of mode identification is equation (20) in D07a.
Here, we write it in a more compact form,
$${\cal A}_k=\varepsilon{\cal F}_k(i). \eqno(1)$$
This equation connects the complex amplitude of the peak,
$${\cal A}_k\equiv A_k\exp(\rm i\varphi_k),\eqno(2)$$
in the $k$-band to the complex amplitude of the associated mode, $\varepsilon$.
This amplitude is related to the
relative surface radius variation through the expression
$${\delta R\over R}=\varepsilon\Theta_\ell^m(\theta)\exp[{\rm i}(m\phi-2\pi\nu t)].$$
The ${\cal F}_k$-coefficients split into two terms,
$${\cal F}_k(i)={\cal D}^k(i)f+{\cal E}^k(i).\eqno(3)$$

The first term arises from the temperature perturbation and the second from the
displacement of the surface element. The $f$-factor describing the response of the flux
to the radius change is a complex quantity which may be obtained from nonadiabatic
calculations together with the instability parameter $\eta$ for a specified stellar
model. The expressions for ${\cal D}^k$ and ${\cal E}^k$, which come from the
integration of local contributions over the stellar disc with the limb darkening factor
and its perturbation taken into account, were given in D07a. The expressions are very
long and need not be repeated here.

Data on stellar surface parameters and atmospheric models are needed to convert changes
in radius and effective temperature into changes in monochromatic fluxes. We need also
information about the rotation rate and the inclination angle. Rotation affects the stellar
model but also, which is more important, the values of the terms ${\cal D}^k$ and ${\cal
E}^k$ through its influence on the $\Theta_\ell^m(\theta)$ dependence.

Given ${\cal F}_k$ and the data from the three passbands, we have three complex linear
equations for $\varepsilon$, implying four degrees of freedom. Thus we may use
$$\chi^2=\frac1{4} \sum_{k=1}^3 w_k |\varepsilon{\cal F}_k(i) -{\cal A}_k|^2, \eqno(4)$$
as the mode and the inclination discriminant, where $w_k$ are
the statistical weights defined by the observational errors
$$w_k=|\sigma_k|^{-2}.$$
For a given pulsation mode and specified stellar parameters, we search for
the best value of $\varepsilon$ to fit photometric amplitudes and phases in
all passbands simultaneously
$$\frac{\partial\chi^2}{\partial\varepsilon}=0.$$
Thus the $\chi^2$-minimization yields the following expression for
the mode's complex amplitude
$$\varepsilon=\frac{\sum_{k=1}^3 w_k {\cal A}_k{\cal F}_k^*}{\sum_{k=1}^3 w_k |{\cal F}_k|^2},\eqno(5)$$
where ${\cal F}_k^*$ is the complex conjugate.

Mode amplitudes, $\varepsilon$, are observational characteristics
that cannot be compared with calculations because nonlinear modeling
of nonradial stellar oscillations is far ahead of us. Nonetheless,
the values of $|\varepsilon|$ deduced from data are of
interest as a hint towards understanding the mechanism of
amplitude limitation. In the present work, the values of
$\varepsilon$ were used to constrain mode identification.

Even if $\chi^2$ is small, we cannot accept identifications if they
imply too large values of $\varepsilon$ because all known SPB stars are low
amplitude pulsators. De Cat (2007), who compiled data on nearly two
hundred (including candidates) SPB stars, found no object with an
amplitude above $A_V=0.03$ mag.  We may use this information to assess
the upper limit of $\varepsilon$ (denoted $\varepsilon_{\rm max}$),
assuming, which seems justified, that the peaks with the highest
amplitudes are due to modes with the highest values of $\varepsilon$ and the
highest values of ${\cal F}_k$. For given stellar model, the highest values
of $|{\cal F}_k|$ occur  at $\ell=1, m=0$ and $i=0^\circ$.
Considering values of $|f|$ (see Eq.\,3) for unstable dipolar modes
in stellar models corresponding to the highest amplitude SPB star HD181558 ($A_V=0.028$ mag) (de Cat, 2007),
we found the average value $\overline{|{\cal F}_V|}\approx$3 mag for $i=0^\circ$.
Therefore, we decided to reject the identification if
$$|\varepsilon|\ge\varepsilon_{\rm max}=0.01.\eqno(6)$$
An additional constraint on $\varepsilon$ follows from the limit of
the radial velocity amplitude of 2 km/s found by J2013. Having the mode
identification and $\varepsilon$ from photometry, we may calculate
the radial velocity amplitude,
 $A_{V_{\rm rad}}^{\rm cal}$, and confront it it
with the observational bound. Equation (18) of D07a yields
$$A_{V_{\rm rad}}^{\rm cal}=2\pi R\nu|{\varepsilon\cal C}_{\rm puls}+{\cal C}_{\rm rot}|,\eqno(7)$$
where ${\cal C}_{\rm puls}$ and ${\cal C}_{\rm rot}$ describe direct
contributions from pulsation and from pulsational modulation of
rotation, respectively. The complicated expressions for these two coefficients
are not required here. It suffices to know that they are fully determined
by the atmospheric parameters of the star and the Hough functions of the mode.

With $\varepsilon$ given by equation\,(5), we get from equation\,(4) a more
explicit expression for our mode discriminant,
$$\chi^2=\frac1{4} \left( \sum_{k=1}^3 w_k|{\cal A}_k|^2
-\frac{\sum_{k=1}^3 w_k|{\cal A}_k{\cal F}_k^*|^2}{\sum_{k=1}^3 w_k
|{\cal F}_k|^2} \right).\eqno(8)$$
Here, unlike the case of the most often used mode discriminants, the data from the three
passbands are treated equally. We see this as the main advantage of our approach to mode
identification.

\subsection{A constraint on the rotation rate}
In the way described above, we may find acceptable mode identification(s) for individual
peaks at a specified confidence level, stellar model and the inclination of the rotation
axis. The range of parameters where a particular identification applies is, in general,
different for different peaks. The requirement of consistency sets constrains on
possible identifications and on admissible ranges of stellar parameters.

The most important are the constraints on rotation. Firstly, because from spectroscopy
we may assess only the value of $V_{\rm rot}\sin i$. Secondly, because some
identifications are acceptable only in very narrow ranges of the inclination.

\subsection{Results}

\begin{figure}
\centering
\includegraphics[width=\columnwidth, clip]{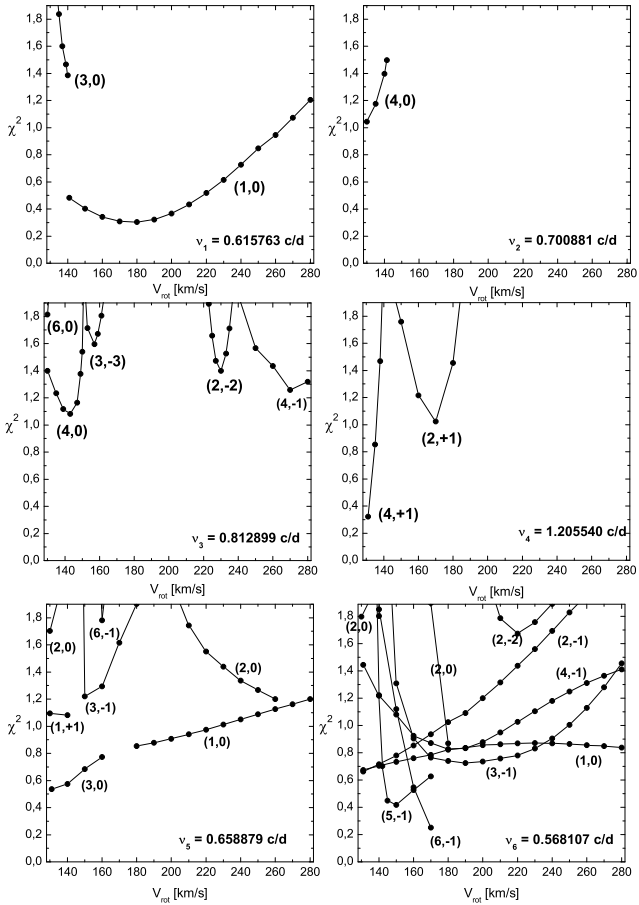}
\caption{Fitting of {\bf eigenmodes} of Model\,1 (see Table\,1) to complex amplitudes for
the six significant frequency peaks in $\mu$ Eri. The dots show minimum values of
$\chi^2$ calculated with equation\,(8). The numbers give the degrees, $\ell$,
and azimuthal orders, $m$, of the modes. The range of $V_{\rm rot}$ corresponds
to a $[90^{\circ},28^{\circ}]$ range of the inclination angle.}
\end{figure}
\begin{figure}
\includegraphics[width=\columnwidth,clip]{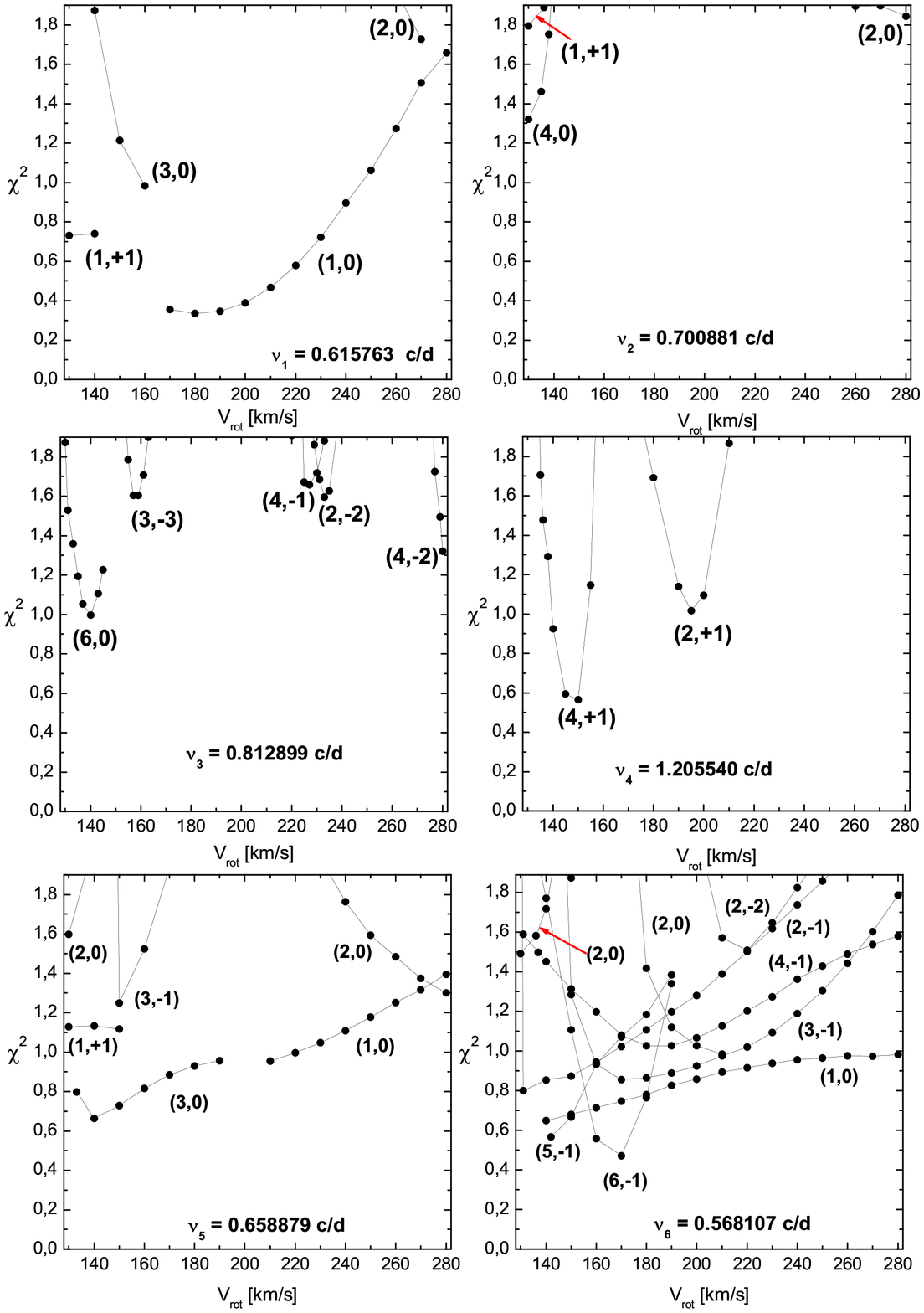}
\caption{The same as Fig.\, 7 but for Model\,2.}
\end{figure}
\begin{figure}
\centering
\includegraphics[width=\columnwidth,clip]{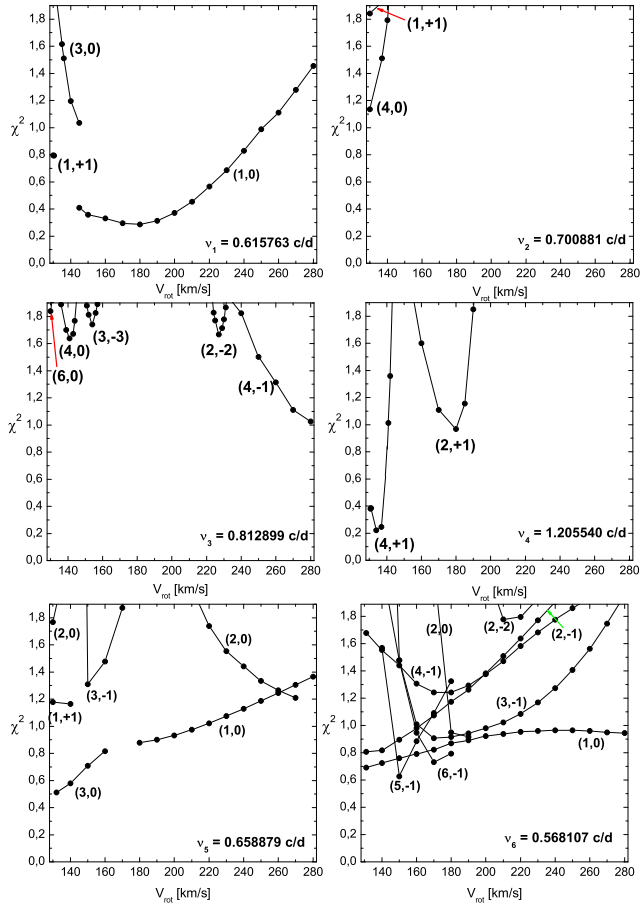}
\caption{The same as Fig.\, 7 but for Model\,3.}
\end{figure}
We adopted \citet{K04} tables based on his stellar atmosphere models for calculations of
${\cal D}^k$,  ${\cal E}^k$,  ${\cal C}_{\rm puls}$ and ${\cal C}_{\rm rot}$ needed in
equations\,(3) and (7). The limb darkening coefficient and its changes were calculated
with the help of \citet{C00} expressions.

Both the rotation rate and inclination of the rotation axis, have significant impact on
mode identification. In order to limit the number of adjustable parameters, we fixed
$V_{\rm rot} \sin i$ at the spectroscopic value of 130 km/s. We did not consider  $\sin
i<0.46$ implying $V_{\rm rot}>280$ km/s because it would be too high for our simplified
treatment. An 80 percent confidence level has been adopted for rejection of a mode
identification. This corresponds to a $\chi^2 ={}$1.5 upper limit for the acceptable
identifications.

The set of modes considered for the six frequency peaks in $\mu$ Eri
included all unstable g-modes with the $\ell$ degrees up to 6 and
r-modes with the $|m|$ orders up to 3. Unstable r-modes with $|m|>3$
have frequencies much higher than the highest frequencies observed
in $\mu$ Eri.  We will discuss the consequences of the bound on
radial velocity amplitude later on but in our initial mode
identification we will take into account only the conditions
$$\chi^2\le1.5, \qquad\varepsilon\le0.01,\qquad\mbox{and}\quad \eta\ge-0.05.\eqno(9)$$

The last condition is a somewhat liberalized (in view of
uncertainties) instability condition.

  As we mentioned in Section\,2, the frequency $\nu_3$ is equal to
$6/P_{\rm orb}$. Therefore, for identification of $(\ell,m)$ of
$\nu_3$ we have to allow also stable modes, i.e., not excited by the
$\kappa$ mechanism. However, to account for the amplitudes,
$\varepsilon$, listed in Tables\,4 to 6, a close proximity of the
stellar eigenfrequency to $6/P_{\rm orb}$ is needed because for the
equilibrium (non-resonant) tide the corresponding amplitudes are by
orders of magnitude lower.

The values of $\chi^2$ for the six peaks plotted against the rotational
velocity are depicted in Figs.\,7, 8, and 9. They were obtained with the use of
Models\,1, 2, and 3, respectively.

As we may see in Fig.\,7, there are two possible identifications for the  $\nu_1$ peak.
The $\ell ={}$1, $m= {}$0 identification is acceptable in the whole range of $V_{\rm
rot}$ we consider, beginning from about 140 km/s. At lower $V_{\rm rot}$,  this peak is
outside the instability range for the $\ell ={}$1, $m ={}$0 modes (see Fig.\,3). Below,
but only down to 138 km/s, the (3,0) identification becomes acceptable and, in fact, the
only acceptable one with the constraint on rotation set by the unique (4,0)
identification for the $\nu_2$ peak. In the very narrow $V_{\rm rot}$ range allowed by
the identification of the $\nu_2$ peak, the unique possibilities are: (4,0) for the
$\nu_3$ peak and (4,+1) for the $\nu_4$ peak. The latter identification imposes an upper
limit on the equatorial velocity of 138.2 km/s and the allowed range of $V_{\rm rot}$
reduces to a small fraction of 1 km/s. Still, for the remaining two peaks we have a
number of options. All acceptable identifications for the six peaks together with the
associated values of $|\varepsilon|$ and $A_{V_{\rm rad}}^{\rm cal}$ are listed in
Table\,3.

\begin{table}
\centering \caption{Modes in Model\,1 that may be associated with
the six frequency peaks in $\mu$ Eri with the constraints given in
equation\,9. The amplitudes $|\varepsilon|$ and $A_{V_{\rm
rad}}^{\rm cal}$ were calculated with the use of equations\,5 and 7,
respectively. The equatorial velocity  $V_{\rm rot}\approx{}$138
km/s, which corresponds to $i\approx{}$70$^{\circ}$, results from
fitting the frequencies $\nu_1$ and $\nu_4$. The corresponding value
of the rotational frequency is $\nu_{\rm rot}=0.44$ c/d.}
\begin{tabular}{|c|c|c|c|c|c|c|c|c|}
\hline
frequency [c/d] & $(\ell,m)$ &   $|\varepsilon|$ & $A_{V_{\rm rad}}^{\rm cal}$ [km/s] \\
\hline \hline
 $\nu_1 ={}$0.615763 & $(3, 0)$ & 0.0055 & 6.0 \\
\hline
 $\nu_2 ={}$0.700881 & $(4, 0)$ & 0.0024 & 0.3 \\
\hline
 $\nu_3 ={}$0.812899 & $(4, 0)$ & 0.0022 & 0.4  \\
\hline
 $\nu_4 ={}$1.205540 & $(4,+1)$ & 0.0032 & 2.1 \\
\hline
 $\nu_5 ={}$0.658879 & $(1,+1)$ & 0.0003 & 1.3  \\
                  & $(3, 0)$ & 0.0030 & 2.2  \\
\hline
 $\nu_6 ={}$0.568107 & $(1, 0)$ & 0.0017 & 1.5 \\
                  & $(2,-1)$ & 0.0015 & 0.7 \\
                  & $(4,-1)$ & 0.0018 & 0.6 \\
\hline
\end{tabular}
\end{table}

The problem with the mode identifications presented in Table\,3 is
caused by the high radial velocity amplitudes that are implied for
some of the peaks. In particular, for the dominant $\nu_1$ peak, the
calculated amplitude exceeds by a factor of 3 the observed limit of
2 km/s. A reliable comparison requires that the spectroscopic value
is derived from the changes in the first moments of spectral lines
and that spectroscopic and photometric data are roughly
contemporaneous. These conditions are not satisfied in our case.

It is important to check how the mode identification depends on the choice  of the
reference model. What matters is not only the change in the instability ranges,
discussed already in Section\,4, but also the modification in mode visibility. The
$f$-factor in equation\,(3) depends on $T_{\rm eff}$ and $L$. Also other terms in this
expression would be somewhat modified by changes of $\lambda$ at fixed $V_{\rm rot}$ and
$\nu_*$.

Comparing  Figs.\,8 and 9 with Fig.\,7 we first note similarities. For instance, always
the (1,0) identification is valid for the $\nu_1$ peak over the wide range of $V_{\rm
rot}$ and (4,0) is the only identification for the $\nu_2$ peak and only in a very
narrow range. However, with the requirement of the same $V_{\rm rot}$ for all peaks, the
acceptable ranges become very narrow and we get significant differences in mode
assignment. The ranges are not much different, $V_{\rm rot}\approx{}$136 km/s with
Model\,2 and 137 km/s with Model\,3. In Tables\,4 and 5, we list the mode
identifications and the corresponding amplitudes. Note, in particular, the differences
for the $\nu_1$ peak, which was assigned to an $\ell ={}$3, $m ={}$0 mode in Model\,1. It should
be noted that for the dominant peak with all models the lowest value of $\chi^2$ was
obtained for the (1,0) identification but outside the $V_{\rm rot}$ range allowed by the
peaks $\nu_2$ and $\nu_4$.

With Model\,3 we obtained almost the same identifications as with
Model\,1. Only one possible identification for the $\nu_6$ peak has
been eliminated. The same identification for the $\nu_1$ peak leads
to even higher $A_{V_{\rm rad}}^{\rm cal}$. With Model\,3, the
problem is the fit for the $\nu_3$ peak which requires rising
$\chi^2$ by about 0.1 above the limit set by conditions\,(9).

With Model\,2, we obtained for the $\nu_1$ peak the (1,+1)
identification and $A_{V_{\rm rad}}^{\rm cal} ={}$3.3 km/s, which is
still above the spectroscopic limit but much less so than in the
other models. In view of the uncertainty in the
spectroscopic limit, the 65\% excess for the $\nu_1$ peak cannot be
regarded as large. It could be further reduced by considering cooler
models, somewhat outside the error box shown in Fig.\,1. The
associated $\varepsilon$ is more than an order of magnitude lower than 0.01.
For all the remaining peaks, except for one of two
possible identifications of $\nu_5$, the calculated radial velocity
is below the 2 km/s limit and $\varepsilon$ is at most 0.006.
\begin{table}
\centering \caption{The same as in Table\,3, but for Model\,2. The
equatorial velocity is $\approx{}$136 km/s, as fixed by fitting the
$\nu_2$ and $\nu_4$ peaks. The corresponding value of the rotational
frequency is $\nu_{\rm rot}=0.42$ c/d.}
\begin{tabular}{|c|c|c|c|c|c|c|c|c|}
\hline
frequency [c/d] & $(\ell,m)$ &   $|\varepsilon|$ & $A_{V_{\rm rad}}^{\rm cal}$ [km/s] \\
\hline \hline
 $\nu_1 ={}$0.615763 & $(1,+1)$ & 0.0008 & 3.3 \\
\hline
 $\nu_2 ={}$0.700881 & $(4, 0)$ & 0.0031 & 0.6 \\
\hline
 $\nu_3 ={}$0.812899 & $(6, 0)$ & 0.0040 & 1.1  \\
\hline
 $\nu_4 ={}$1.205540 & $(4,+1)$ & 0.0033 & 1.3 \\
\hline
 $\nu_5 ={}$0.658879 & $(3, 0)$ & 0.0056 & 2.1  \\
                  & $(1,+1)$ & 0.0005 & 1.5 \\
\hline
 $\nu_6 ={}$0.568107 & $(2,-1)$ & 0.0016 & 0.7  \\
                  & $(4,-1)$ & 0.0019 & 0.5  \\
\hline
\end{tabular}
\end{table}
\begin{table}
\centering \caption{The same as in Table\,3, but for Model\,3. The
rotation velocity is $\approx{}$137 km/s as fixed by fitting the
$\nu_1$ and $\nu_2$ peaks. The corresponding value of the rotational
frequency is $\nu_{\rm rot}=0.44$ c/d. In order to fit an
$\ell={}$4, $m ={}$0 mode to $\nu_3$, we had to increase $\chi^2$ by
about 0.2 above the limit set in equation\,(9).}
\begin{tabular}{|c|c|c|c|c|c|c|c|c|}
\hline
frequency [c/d] & $(\ell,m)$ &   $|\varepsilon|$ & $A_{V_{\rm rad}}^{\rm cal}$ [km/s] \\
\hline \hline
 $\nu_1 ={}$0.615763 & $(3, 0)$ & 0.0066 & 6.4 \\
\hline
 $\nu_2 ={}$0.700881 & $(4, 0)$ & 0.0028 & 0.1  \\
\hline
 $\nu_3 ={}$0.812899 & $(4, 0)$ & 0.0024 & 0.6  \\
\hline
 $\nu_4 ={}$1.205540 & $(4,+1)$ & 0.0037 & 2.2  \\
\hline
 $\nu_5 ={}$0.658879 & $(3, 0)$ & 0.0037 & 2.4 \\
                  & $(1,+1)$ & 0.0003 & 1.3 \\
\hline
 $\nu_6 ={}$0.568107 & $(1, 0)$ & 0.0021 & 1.6 \\
                  & $(2,-1)$ & 0.0016 & 0.7  \\
\hline
\end{tabular}
\end{table}

The modes we associated with $\nu_3$ are, in fact, unstable but
for the tidal excitation this is irrelevant. What matters is only
that the driving or damping rate is much lower than the frequency.
However, it is important to stress that we end up with the same
identification of the $\nu_3$ peak considering only unstable modes.
None of the $|\varepsilon|$ values listed in Tables\,3
- 5 comes close to 0.01, which was set as the upper limit for
acceptable identifications.

We cannot  exclude very different identifications if the equatorial
velocity is above 280 km/s, which we adopted as the maximum only
because of limitations of the method we used. It is fortunate that
consistent identifications were found well below this value. With
$V_{\rm rot}<140$ km/s, the star rotates at less than 1/3 of the
maximum rotation rate and the excess of the equatorial radius over
the mean radius is below 4 percent. The stabilization of retrograde
modes was found by \citet{L10} much closer to maximum equatorial
velocity. Thus, there should be no reservation against our few
identification with such modes.

\section{Summary and conclusions}

Using three-band photometric data, we determined angular numbers,
$\ell$ and $m$, of modes responsible for the six significant
frequency peaks in the fast rotating SPB star $\mu$ Eri.
Simultaneously, narrow ranges for the allowed equatorial velocity of
rotation were found. We took into account modes with $\ell\le6$ in
three models with mean parameters strongly constrained by
observational data. The modes were calculated with a nonadiabatic
version of {\it the traditional approximation}. For the five
frequency peaks only unstable modes were taken into account whereas
for the one frequency $\nu_3=6/P_{\rm orb}$ this condition was
relaxed. Nevertheless, the identified modes for this frequency
turned out to be unstable.
Thus, we cannot decide whether the $\nu_3$ peak arises from the
tidal action exerted by the companion or from excitation by the
$\kappa$-mechanism operating in the $Z$-bump zone. Fortunately, this
uncertainty neither affects constraints on the rotation rate nor the
mode identification. The inferred equatorial velocity is slightly
different for the three models but for all of them it is between 135
and 140 km/s.

One interesting difference between the models was found for the dominant peak, which was
attributed to a zonal $\ell ={}$3 mode in two models (Models\,1 and 3) and to a prograde
$\ell ={}$1 mode in one (Model\,2). These different identifications were found with Models\,1
and 2 lying on the same $6 ~{}$M$_{\sun}$ evolutionary track and differing by less than 0.008
in $\log T_{\rm eff}$. With Model\,3 ($M= {}$5.85~M$_{\sun}$, $\log T_{\rm eff}$ nearly the
same as Model\,2) all angular numbers were found to be the same as in Model\,1, except
for one alternative identification for $\nu_3$. With all our models, only the $\ell ={}$4
and $m ={}$0 numbers and only at the lowest equatorial velocity allowed by $V_{\rm rot}\sin
i$ data  were found acceptable for the second in the significance hierarchy peak
($\nu_2$). This eliminated the $\ell ={}$1 and $m ={}$0 possibility for the dominant peak
which otherwise would have been acceptable in the wide range of $V_{\rm rot}$ for all
three models.

With known $V_{\rm rot}$, $\ell$, and $m$, we calculated amplitudes of the radial
velocity variations and compared them with spectroscopic data, which yielded only an upper bound.
The $\ell ={}$3, $m ={}$0 identification for $\nu_1$ leads to the amplitude exceeding the upper
bound by at least a factor of three. With $\ell ={}$1, $m ={}$1, the excess is much lower. It
would be going too far to judge that Model\,2 is better than the remaining two models
because the spectroscopic and photometric data do not satisfy criteria of a reliable
comparison, as specified in Section\,5.3. Nonetheless, the large difference in the
calculated amplitudes points to the prospect for constraining model parameters with
better data on the radial velocity amplitudes.

There are some differences in $\ell$ inferred with Model\,2 and with the remaining two
models. For the $\nu_3$ peak, $m ={}$0 is found with all models but $\ell ={}$4 is inferred
with Models\,1 and 3 while $\ell ={}$6 with Model\,2. For the $\nu_2$ peak the $\ell ={}$4,
$m ={}$0 identification is found with all the models whereas for the $\nu_4$ peak we got
$\ell ={}$4, $m ={}+$1. For the lowest amplitude peaks there are alternative identifications.

The inferred degrees range from $\ell ={}$1 to 6. It seems that there is no way to predict
which of the huge number of unstable modes will reach the amplitude at a specified
detection level. Neither driving rate nor visibility is a reliable predictor. Perhaps
the preference to the $\ell ={}$4 modes results from a compromise between visibility and
the number of unstable modes in the frequency range of the peaks. This number is much
higher at $\ell ={}$4 than 1. Pulsations in SPB stars appear less predictable than the
stochastic solar-like oscillations.

Without constraints on the angular numbers from multiband photometry
the task of mode identification in SPB stars appears totally
hopeless. In Section\,1, we stressed that an equidistant separations
in period found by CoRoT are most likely misleading. The $\ell\le2$
limit on the potential identifications is not justified. In the
present work, we considered modes up to $\ell ={}$6 but it may still
be not enough.

The degree of excited modes even at the moderate rotation rates may only be determined
jointly with azimuthal orders and the inclination of the rotation axis. This complicates
the mode identification but as a reward we obtain valuable information about
rotation. The result is sensitive to model parameters. Adequate data on the radial
velocity amplitude lead to a new constraint on the model. Unfortunately, for $\mu$ Eri
we have only upper limits. This is one of the sources of uncertainties of our results.
Data on radial velocity amplitudes are very important but to be fully useful they have
to come from an epoch not too distant from that of the photometric data and must be based on
measurements of changes in the first moments of spectral lines which measure
disc-averaged velocities.

\section*{Acknowledgments}
 We are indebted to the anonymous referee for pointing out that the
frequency $\nu_3$ is a whole multiple of the orbital frequency. The
work of JDD and GH was supported by the Polish NCN grants
2011/01/M/ST9/05914 and 2011/01/B/ST9/05448. WD was supported by the
Polish NCN grant DEC-2012/05/B/ST9/03932.

\end{document}